# The relationship between R&D spillovers and regional innovation: Licensing patents through royalties and the Stackelberg duopoly with subgame perfect Nash equilibrium

Vasilios Kanellopoulos

University of Birmingham

School of History and Cultures

Edgbaston, Birmingham B15 2TT

United Kingdom

∗Corresponding Author: Vasilios Kanellopoulos e-mail: kanellopbill@gmail.com

**Abstract:** The present paper examines the effect of R&D spillovers on regional innovation in Greece over the 2002-2010 period. The approach taken goes beyond a regional knowledge production function and draws possible explanations from a more extensive pool of R&D related and regional structural variables. Having employed game theory techniques in order to describe the licensing of the patents through royalties and derived the subgame perfect Nash equilibrium under a Stackelberg duopoly, the results obtained accord with findings of previous studies when it comes R&D expenditure related variables and further suggest that the role of highly-qualified employment is instrumental in promoting regional innovation. The results also suggest the benefits of synergies between R&D personnel in manufacturing and other measures of highly-qualified employment as well as R&D expenditure of the public sector and employment in manufacturing business R&D for regional innovation.

## 1. Introduction

This research examines the effect of research and development (R&D) spillovers on regional innovation in Greece over the 2002-2010 period.

The idea of R&D spillovers and their effects for the regional economies have been seen as a promising research avenue (Ghio et al. 2015). Much of the existing literature has concentrated on the importance of knowledge and R&D spillovers for regional economic growth (Sterlacchini 2008; Varga and Schalk 2004), while others have focused their analysis at the firm level and have examined the role of R&D in innovation activity. The present research finds its place within the literature exploring the effect of R&D spillovers on regional innovation. Within the context of this literature, most empirical studies regarding the influence of R&D spillovers on regional innovation take place either in a U.S. context (Acs et al. 1992, 1994a; Anselin et al. 1997; Jaffe 1989) or the prosperous economies of Europe (Del Barrio-Castro and Garcia-Quevedo 2005; Fischer and Varga 2003; Fritsch and Franke 2004; Piergiovanni and Santarelli 2001). These studies find a positive effect of R&D spillovers on regional innovation.

This research thus aspires to shed some light on the relation between R&D spillovers and regional innovation in the Greek context. In doing so, the paper deploys an array of explanatory variables that go beyond the basic R&D indicators usually employed in the literature. Within this augmented array of explanatory variables, high technology employment related variables, such as the share of employment in high-tech manufacturing and services and the intensity of scientists and engineers that have been suggested by some authors as Audretsch and Feldman (1996) and Feldman (1999), have a prominent place.

In addition, the paper assesses the differences in the effect size of different sector-related R&D spillovers measures. In this instance, the results produced in the relative literature have been deemed controversial depending on the country context. A number of studies have produced innovation elasticities for the private sector R&D which are higher than the corresponding elasticities for university R&D related measures (Autant-Bernard 2001; Jaffe 1989; Piergiovanni et al. 1997; Ronde and Hussler 2005). This contrasts the directly opposing findings in some other studies (Acs et al. 1992; Blind and Grupp 1999; Del Barrio-Castro and Garcia-Quevedo 2005). The present research enters this discussion while also placing more emphasis on high technology employment-related measures.

Another interesting point of our analysis is that it accounts for possible synergies of R&D spillovers originating from different sources and examines their effect on regional innovation. This succeeds it by employing interaction terms between different sources (indicators) of R&D spillovers. The relative literature regarding the effect of the collaboration between different R&D actors is relatively small (Charlot et al. 2015; Fritsch and Franke 2004; Ponds et al. 2010).

However, the point of the present research which draws the attention and makes it original is that it employs sophisticated methodology. More specifically, it introduces us to the game theory and assumes a Stackelberg duopoly model where the leader and the follower produce their quantities. At the same time, the leader produces a superior technology and as a patentee seeks to determine a license fee or/and royalty to charge the follower-licensee for the license to use the patent. He advances to this action in order to maximize his/her profit. As a result, in this model, the patentee is the leader in setting a license fee or/and royalty and he/she is also the leader in producing quantities (production of the leader's quantity in the Stackelberg model). This point is the novel characteristic and main difference of the present research compared to previous studies which treat the patentee as an outsider (he does not operate in the industry). On the other hand, the follower- licensee competes in the product market under the Stackelberg duopoly model and he/she is willing to pay the fee or and royalty (price of license) in order to use the patent. All the information regarding the introduction of the patent licensing through royalties, as in our case, enters the Stackelberg duopoly model and the subgame perfect Nash equilibrium is then determined.

The results obtained by this study suggest that all the variables of R&D spillovers relating to both R&D activities and skilled labor have a strongly positive and significant effect on regional innovation. Moreover, when different types of R&D spillovers are compared, it is the estimated elasticity of patents in respect to the intensity of scientists and engineers as well as to the share of employment in high-tech manufacturing and services that are higher than the patent estimated elasticities for the other R&D spillovers measures. This is an interesting finding that highlights the importance of people rather than the expenditure, something that may have been overlooked in other studies.

The results on interactions between spillovers originating from different sources suggest that synergies may exist between these sources that together shape up a more conducive to innovation 'climate' at the regional level.

When it comes to the effect of the other variables deployed, proxies of regional economic diversity and agglomeration economies were found to be important drivers of regional innovation activity. Furthermore, GDP growth has a positive and significant effect on regional patent intensity whereas the effect of unemployment was insignificant. In contrast regional specialization was found to have a negative effect on regional innovation alongside the effects of both the manufacturing enterprise's size and the local share of services.

The present paper is organized as follows: the next section presents the theory, while the third presents the methodology. The fourth section resorts to a detailed description of the data and variables. Section five describes the econometric methods used and the empirical results. Section six, finally, contains conclusions and policy implications.

## 2. Theory

*2.1 The concept of R&D spillovers*

Alternative definitions of R&D spillovers exist in the literature. Griliches (1992: p. S36-S37) defines R&D spillovers as "working on similar things and hence benefitting much from each other's research". According to Agarwal et al. (2010: p. 271), knowledge spillovers are defined as "the external benefits from the creation of knowledge that accrue to parties other than the creator, occur at multiple levels of analysis, be it within or across organizations and networks". Central to the theorization of knowledge spillovers is the public good character of knowledge (Agarwal et al. 2007; Arrow, 1962). Arrow (1962) introduces the concepts of non-rivalness and non-excludability. The non-rivalness concept supports that the use of knowledge by an economic agent does not preclude others from the use of it, whereas the non-excludability, in turn, pertains to the fact that the owner of an innovation or knowledge cannot prevent others from accessing and using it. These public characteristics of knowledge allow its transfer and absorption from third firms (Agarwal et al. 2007; Agarwal et al. 2010).

*2.2 The effect of R&D spillovers on regional innovation*

The empirical evidence accumulated suggests that there are positive R&D spillovers (both industrial and university) effects on innovation (Acs et al. 1992, 1994a; Anselin et al. 1997; Jaffe 1989) and spatial concentration of innovation (Audretsch and Feldman 1996) at a U.S. state level. Apart from R&D expenditures, high-tech labor has been also suggested as a measure of R&D spillovers and evidence has been produced for its positive impact on regional innovation (Audretsch and Feldman 1996). Anselin et al. (1997) examine the effect of R&D spillovers at different levels of spatial disaggregation and provide evidence suggesting that spillovers are localized at sub-state levels. University spillovers present a strong positive impact on innovation within both metropolitan statistical areas (MSA) and counties, while industrial R&D spillovers benefits are contained only within the MSA (Anselin et al. 1997). This concurs with the findings of Bottazzi and Peri (2003: p. 687) who maintain that "doubling R&D spending in a region would increase the output of new ideas in other regions within 300 km only by 2–3%, while it would increase the innovation of the region itself by 80-90%". A number of studies have put forward evidence favoring the prominence of intra-regional R&D spillovers on innovation (Autant-Bernard 2001; Del Barrio-Castro and Garcia-Quevedo 2005; Piergiovanni and Santarelli 2001; Ronde and Hussler 2005). However, there have been other studies which provide positive evidence for inter-regional R&D spillovers (Fischer and Varga 2003; Moreno et al. 2005; Crescenzi et al. 2007).

The interaction between different measures of knowledge spillovers and its effect on regional innovation has been the distinguishing feature in a number of studies. Charlot et al. (2015) provide evidence suggesting that the interaction between R&D and human capital is a factor that increases regional innovation and argue that R&D positively affects innovation only in local areas where a significant amount of complementary skilled labor also exists. It is explained that the latter is also able to contribute to the generation of new knowledge and ideas. The empirical results produced in the literature also demonstrate that it is industrial and university collaborations that promote regional innovation (Fritsch and Franke 2004; Ponds et al. 2010; Ronde and Hussler 2005).

## 3. Licensing patents through royalties and the Stackelberg duopoly with subgame perfect Nash equilibrium

We consider a duopoly in the $r_{th}$ region of a country. Thus, there are two firms, $i = 1,2$. There, the firm 1 is the leader and the firm 2 is the follower. As a result, a Stackelberg model takes place. These firms produce a homogeneous good. In this market, the linear inverse demand function is given by:

$$p = a - Q, \qquad (1)$$

where $Q = q_1 + q_2$ is the total quantity of the market, $q_1$ is the quantity produced by the leader, $q_2$ is the quantity produced by the follower and $p$ is the price of the product.

In this market, the leader is also a patentee. Besides the produced quantity described above, he/she also produces a superior technology. Here, the patentee seeks to determine a license fee or royalty that will maximize his/her profit (Kamien and Tauman 1983, 1986). At the same time, the follower firm is a licensee (Kamien and Tauman 1983, 1986; see also Kamien et al. 1992). In more detail, the patentee seeks to license the patent to the follower-licensee[1] in order to maximize his/her profit. In other words, the patentee sets and chooses a license fee or/and royalty to charge the follower-licensee for the license to use the patent[2].

An important point of the present research to be underlined is that the patentee is not an outsider and, thus, operates in the industry[3]. As a result, the patentee is the leader in setting a license fee or/and royalty[4] (royalty in our case) and he/she is also the leader in producing quantities (production of the leader's quantity in the Stackelberg model). This is the main difference of our research compared to previous studies. On the other hand, the follower produces its quantity in the duopoly Stackelberg model and he/she is also a licensee as he/she is willing to pay the fee or and royalty (price of license) in order to use the patent (see Kamien et al. 1992)[5]. Given that, in the Stackelberg model, firms decide sequentially on the output they produce[6], to find the Nash equilibrium of this sequential game, we use backward induction. That is, we first solve the optimization problem for the follower, and with this information determine the optimal choice by the leader.

---

[1] In our case, we have only licensees who purchase the license. In Kamien and Tauman (1983, 1986), the firms decide whether or not to purchase the license.

[2] In our analysis, we focus on the licensing by means of a royalty. Here, the patentee chooses a royalty $r$ ($r > 0$) which a licensee is required to pay per unit of production.

[3] Kamien and Tauman (1986) assume that the patentee is an outsider and does not operate in the industry.

[4] Kamien and Tauman (1986: p. 472) note that "the inventor plays the role of a Stackelberg leader in the game as he determines the reaction function of the license price, or the demand function for licenses, and maximizes against it". The same view is also supported by Kamien et al. (1988).

[5] In their analysis, Kamien et al. (1992: p. 484) support that "the patentee is treated as a leader in a Stackelberg type game in which the potential licensees, the followers, compete in the product market".

[6] First, the leader firm chooses its production quantity. This decision cannot be changed after. Sequentially, the follower firm chooses its output after observing the quantity chosen by the leader. This is the key difference when compared Cournot model, in which the production decisions by the firms are taken simultaneously.

As a result, the follower firm which is simultaneously a licensee (purchasing the license) chooses $q_2$, taking the quantity produced by the leader into consideration. Thus, to derive the optimal output, we need to find the quantity $q_2$ that maximizes the profit function for the follower firm, taking $q_1$ (the quantity of the leader) as given. Then, the profit of the follower is:

$$\pi_2(r, q_1, q_2) = p(q_1 + q_2)q_2 - r^2 q_2 - c q_2, \tag{2}$$

or

$$\pi_2(r, q_1, q_2) = (a - Q)q_2 - r^2 q_2 - c q_2, \tag{3}$$

or

$$\pi_2(r, q_1, q_2) = \big(a - (q_1 + q_2)\big)q_2 - r^2 q_2 - c q_2, \tag{4}$$

or

$$\pi_2(r, q_1, q_2) = a q_2 - q_2(q_1 + q_2) - r^2 q_2 - c q_2, \tag{5}$$

or

$$\pi_2(r, q_1, q_2) = a q_2 - q_2 q_1 - q_2^2 - r^2 q_2 - c q_2, \tag{6}$$

Here, $r$ is the royalty (the price of the license)[7], $q_1$ is the quantity of the leader, $q_2$ is the quantity of the follower and $c$ is the constant marginal cost, $c > 0$.

Then, the maximization problem resolves to:

$$\frac{\partial \pi_2}{\partial q_2} = a - q_1 - 2q_2 - r^2 - c = 0, \tag{7}$$

---

[7] In our research, the license is sold by means of a non-linear royalty. In other studies, license is sold by means of a linear royalty (see Kamien and Tauman 1983, 1986; Kamien et al. 1992).

or

$$2q_2 = a - q_1 - r^2 - c, \tag{8}$$

or

$$q_2^* = \frac{a-q_1-r^2-c}{2}, \tag{9}$$

The equation (9) is the reaction function of the follower. This is the best response function of the follower to any quantity produced by the leader.

On the other hand, the leader chooses $q_1$ with the understanding that the follower will react to this choice according to its reaction function described in the equation (9). To find the optimal quantity produced by the leader, we introduce the optimal response (best response) of the follower into the leader's profit function in place of $q_2$. Then, the leader's profit function is:

$$\pi_1(r, q_1, q_2) = p(q_1 + q_2)q_1 + r^2 q_1 - cq_1, \tag{10}$$

or

$$\pi_1(r, q_1, q_2) = (a - Q)q_1 + r^2 q_1 - cq_1, \tag{11}$$

or

$$\pi_1(r, q_1, q_2) = \big(a - (q_1 + q_2)\big)q_1 + r^2 q_1 - cq_1, \tag{12}$$

or

$$\pi_1(r, q_1, q_2) = \big(a - (q_1 + q_2(q_1))\big)q_1 + r^2 q_1 - cq_1, \tag{13}$$

Substituting for $q_2(q_1)$ from the follower's problem, we have:

$$\pi_1(r, q_1, q_2) = \left(a - \left(q_1 + \frac{a-q_1-r^2-c}{2}\right)\right)q_1 + r^2 q_1 - cq_1, \tag{14}$$

or

$$\pi_1(r, q_1, q_2) = aq_1 - q_1^2 - q_1 \frac{a - q_1 - r^2 - c}{2} + r^2 q_1 - c q_1, \quad (15)$$

or

$$\pi_1(r, q_1, q_2) = aq_1 - q_1^2 - q_1 \left(\frac{a}{2} - \frac{q_1}{2} - \frac{r^2}{2} - \frac{c}{2}\right) + r^2 q_1 - c q_1, \quad (16)$$

or

$$\pi_1(r, q_1, q_2) = aq_1 - q_1^2 - \frac{q_1 a}{2} + \frac{q_1^2}{2} + \frac{r^2 q_1}{2} + \frac{c q_1}{2} + r^2 q_1 - c q_1, \quad (17)$$

Then, the maximization problem resolves to:

$$\frac{\partial \pi_1}{\partial q_1} = a - 2q_1 - \frac{a}{2} + q_1 + \frac{r^2}{2} + \frac{c}{2} + r^2 - c = 0, \quad (18)$$

or

$$a - q_1 - \frac{a}{2} + \frac{r^2}{2} + \frac{c}{2} + r^2 - c = 0, \quad (19)$$

or

$$\frac{2a - 2q_1 - a + r^2 + c + 2r^2 - 2c}{2} = 0, \quad (20)$$

or

$$a - 2q_1 + 3r^2 - c = 0, \quad (21)$$

or $\qquad 2q_1 = a + 3r^2 - c, \qquad (22)$

or $\qquad q_1^* = \frac{a + 3r^2 - c}{2}, \qquad (23)$

The equation (23) is the optimal value for the quantity of the leader and, simultaneously, the reaction function of the leader. Substituting $q_1^*$ into the reaction function of the follower, we take:

$$q_2 = \frac{a - \frac{a+3r^2-c}{2} - r^2 - c}{2}, \qquad (24)$$

or

$$q_2 = \frac{\frac{2a-a-3r^2+c-2r^2-2c}{2}}{2}, \qquad (25)$$

or

$$q_2^* = \frac{a-5r^2-c}{4}, \qquad (26)$$

Here, $q_2^*$ is the output produced by the follower in the equilibrium and, thus, the optimal value.

Given that the patentee also seeks to determine and set a royalty that will maximize his/her profit, the maximization problem also resolves to[8]:

$$\frac{\partial \pi_1}{\partial r} = rq_1 + 2rq_1 = 0, \qquad (27)$$

or

$$3rq_1 = 0, \qquad (28)$$

Dividing both sides of the equation (28) by $r$, we have:

$$3q_1 = 0, \qquad (29)$$

---

[8] Kamien and Tauman (1986) and Kamien et al. (1988) describe the optimal license fees for a new product.

Putting the optimal value of $q_1$ ($q_1^*$) found previously into the equation (29), we obtain:

$$3\frac{a+3r^2-c}{2} = 0, \qquad (30)$$

or

$$3(a + 3r^2 - c) = 0, \qquad (31)$$

or

$$3a + 9r^2 - 3c = 0, \qquad (32)$$

or

$$9r^2 = 3c - 3a, \qquad (33)$$

or

$$r^2 = \frac{3(c-a)}{9}, \qquad (34)$$

or

$$r^2 = \frac{c-a}{3}, \qquad (35)$$

or

$$r^* = \sqrt{\frac{c-a}{3}}, \qquad (36)$$

Here, $r^*$ is the optimal value of the price of the license expressed by royalties. As a result, $(q_1^*, q_2^*, r^*)$ constitute the subgame perfect Nash equilibrium in this Stackelberg duopoly.

## 4. Data and variables

The present paper examines the effect of R&D spillovers and other control variables on regional innovation in Greece. The level of spatial aggregation used was dictated by the data available and pertains to thirteen Greek regions (NUTS 2). The time span of the analysis covers the period from 2002-2010 providing 117 observations for the pooled sample.

*4.1 Dependent variable*

The variable of interest and dependent variable of our model is the patent intensity (**PATINT**), which is defined as patent applications divided by population (Carlino et al. 2007) to account for different regional sizes. The patent data used come from Eurostat's regional statistics.

*4.2 Independent variables*

A number of different but complementary proxies for R&D spillovers were used. The first proxy of R&D spillovers is that of manufacturing R&D expenditures density (**RDEXP**). This measure is defined as regional manufacturing R&D expenditures per $km^2$. This has been previously used by Bottazzi and Peri (2003). Information is also available at the regional level for manufacturing sector R&D personnel. In this way, a manufacturing R&D personnel density variable (**RDPERS**) was constructed for each region and year by taking the ratio of regional R&D personnel in manufacturing sector to regional $km^2$. Anselin et al. (1997) construct a similar variable using regional employment in high technology research laboratories for the U.S. The data used in the present research for constructing **RDEXP** and **RDPERS** come from Hellenic Statistical Authority's (EL.STAT.), Annual Industrial Survey.

Apart from manufacturing R&D related variables, two additional variables are used to account for regional R&D expenditure intensity in higher education (**RDHIGHED**) and in the government sector (**RDGOV**). In both cases the corresponding R&D expenditures were divided by regional km$^2$ to form the respective R&D expenditure densities. For the construction of both variables, data from Eurostat's regional statistics were used.[9]

Going beyond manufacturing R&D expenditure and personnel variables, the present research places a strong emphasis on regional employment-related variables which are thought of as being involved in both innovation production and knowledge spillovers. In particular, the following regional employment variables are used: a) scientists and engineers' density (***SCIENGIN:*** number of scientists and engineers/ km$^2$), b) science and technology density (***EMPSCITECH***: employees in science and technology/km$^2$), c) the share of employment in high technology manufacturing and services (***HTMANSERV:*** employees in high technology manufacturing and high technology knowledge intensive services over total regional employees) and d) the share of employment in high technology manufacturing (***HTMAN:*** employees in high and medium high technology manufacturing over total regional employees). Their introduction follows the rationale discussed above and the construction of similar variables by Audretsch and Keilbach (2007) for scientists and engineers, and Bishop (2012) for employment in high technology sectors. The data for the construction of these variables come from Eurostat's regional database.

*4.3 Control variables*

---

[9] Due to the existence of missing values for some of the years considered, an imputation procedure has been applied. The missing values for Government R&D expenditures have been imputed by regionally apportioning the available corresponding national R&D expenditure figures based on regional shares of persons with tertiary education and/or employed in science and technology. The missing values for Higher Education R&D expenditures were imputed by regionally apportioning national Higher Education R&D expenditure figures based on known regional shares of education sector employment. The correlation coefficients between imputed and real R&D expenditure values for the years with no missing observations were very high. Despite these, the results obtained on the coefficients of extensively imputed variables should be treated as tentative.

Feldman and Audretsch (1999) find that regional diversity, by bringing together complementary activities, is more beneficial than regional specialization for regional innovation. A region with more diversified industrial structure is more capable to escape from the cognitive lock-in created by traditional regional specialization and this is something that facilitates regional innovation activity (Boschma 2005). Van der Panne (2004), however, offers some contrasting evidence as regional variety was found to have a negative effect on regional innovation.

The relative literature offers mixed results when it comes to the effect of regional specialization on innovation. Whereas Feldman and Audretsch (1999) and Crescenzi et al. (2007) find a negative effect, Van der Panne (2004) and Marrocu et al. (2013) demonstrate that a positive and significant influence occurs.

Regional diversity was proxied here by using the Theil (*THEIL*) entropy-based diversity index[10] (Theil 1972: p. 26). Specifically, the Theil index is defined as:

$$T = \sum_i \frac{p_{ri}}{p_r} \log \frac{p_r}{p_{ri}}, \qquad (37)$$

where $E_{ri}$ is the employment in region $r = 1, \ldots, R$ and industry $i = 1, \ldots, J$, $p_{ri} = \frac{E_{ri}}{\sum_r \sum_i E_{ri}}$, and $p_r = \sum_i p_{ri}$. This measure takes the value of 0 when only one sector is present in region r and the value $\ln(n)$ where all $n$ two-digit industrial sectors use the same number of persons in the region in question.

---

[10] The present paper employs the concepts of unrelated (*UNRELATED*) and related variety (*RELATED*) concepts (Frenken et al. 2007) as alternative measures of regional diversity. An entropy-based measure of unrelated variety is defined as: $UV = \sum_{g=1}^{G} P_g \log_2 (\frac{1}{P_g})$. In our analysis, all two-digit industries $i$ are part of a one-digit sector $S_g$, for $g = 1, \ldots, G$. Then, one-digit shares, $P_g$, arise from summing up the two-digit employment shares $p_i$ such that: $P_g = \sum_{i \in S_g} p_i$. In turn, related variety is calculated as the weighted sum of entropy within each one-digit sector: $RV = \sum_{g=1}^{G} P_g H_g$, such that $H_g = \sum_{i \in S_g} \frac{P_i}{P_g} \log(\frac{1}{P_i/P_g})$. In order to construct related variety, Frenken et al. (2007) use five-digit sectors $i$ which fall under a two-digit sector $S_g$, where $g = 1, \ldots, G$. Our analysis, however, follows Caragliu et al. (2016) who instead use two-digit sectors $i$ which fall under a one-digit sector $S_g$ ($g = 1, \ldots, G$). This happens because of a lack of data about employment in a more disaggregated level of analysis (e.g. three-digit, four-digit etc.) regarding Greek regions. The sum of related and unrelated variety gives the *THEIL* index.

On the other hand, regional specialization was proxied by using the Hoover specialization index (Hoover 1941). The Hoover index (**HOOVIND**) is defined as:

$$HOOVIND = \frac{1}{2} * \sum_i \left| \frac{E_{ir}}{E_r} - \frac{E_{in}}{E_n} \right|, \tag{38}$$

where $E_{ir}$ is the employment in industry $i = 1, ...., J$ and $r = 1, ...., R$ regions at some point in time. $E_{in}$ is the employment in industry $i$ nationally, while $E_r$ and $E_n$ pertain to total manufacturing employment at the regional and national level, respectively.

The effect of local economic conditions was proxied by GDP growth (**GDPGR**). Here an annual growth rate with one lag was used. In empirical studies, GDP per capita seems to positively affect the regional innovation process (Corradini and De Propris 2015; Moreno et al. 2005). In the former case, this finding is stronger when spillover effects from neighboring regions are taken into account. Moreno et al. (2005) note that per capita GDP is a measure of economic wealth, while Corradini and De Propris (2015) underline that GDP per capita indirectly accounts for the size of the capital stock in the region. Population growth (**POPGR**) enters the model as a proxy of local dynamism. This auxiliary variable enters the model in order to provide a further check for the effect of GDP growth on regional innovation.

Crescenzi et al. (2007: p. 684) support that the "rate of unemployment (and in the case of the EU especially its long-term component) is an indicator of the rigidity of the labor market and of the presence of individuals whose possibilities of being involved in productive work are hampered by inadequate skills". In contrast to GDP growth, the unemployment rate has been found to have a negative effect on regional innovation in different regional contexts[11] (Crescenzi et al. 2007). The unemployment rate was used in the present study with four (*UNEMP4*) and one-year (*UNEMP1*) lags.

Another control variable used in the present paper is the regional employment share of the services sector (**LOCALSERV**). The local presence of the services sector has been seen as being responsible for providing the local economy with important inputs such as counseling, technical services, venture capital and a technological infrastructure that leads to a higher innovation activity (Fritsch and Slavtschev 2011; see also Feldman and Florida 1994). It has, however, also been argued that as manufacturing is characterized by more extensive patenting than services (Bottazzi and Peri 2003; see also Bode 2004; Fritsch and Slavtschev 2010, 2011), the de-industrialization and the overrepresentation of the services sector in a local economy could also mean less innovative activity and hence a negative effect can also be possible (Fritsch and Slavtschev 2011).

Agglomeration economies are proxied by population density (**POPDENS:** regional population per km$^2$). According to Fritsch and Slavtschev (2011), population density describes not only urbanization economies but also diverse types of unobserved region-specific influences. As such, densely populated regions create conducive conditions for interaction and provide a rich physical and institutional infrastructure (Fritsch and Slavtschev 2010, 2011).

Acs et al. (1994b) provide evidence suggesting that while own-R&D expenditure is relatively more important as an input for generating innovation for the large firms, it is the knowledge spillovers from university research that becomes more important in accounting for the innovative activity of the small firms. As small firms are the main beneficiaries of knowledge spillovers, the regional average firm size (**SIZE:** average firm employment size) is also included amongst the control variables. Fritsch and Slavtschev (2010, 2011) demonstrate that average firm size has a negative and significant effect on regional innovation, and similar evidence has been found for U.S. metropolitan statistical areas by Anselin et al. (1997).

---

[11] Crescenzi et al. (2007) explore the effect of unemployment on regional innovation in both U.S. MSAs areas and EU regions.

## 5. Econometric methods and empirical results

In order to identify the relative importance of systematic sources of variation and test for possible regional and time effects in the dependent variable as well as key explanatory variables. Table 1 presents the results of this variance decomposition analysis and F-tests for the existence of possible regional and time effects. The dependent variable in the econometric analysis to follow (**PATINT**) is characterized by significant between-regions variance that accounts for 64% percent of the overall variation whereas the time variance amounts to only 4% of the overall variation. The relative time invariance is even more pronounced in the case of key explanatory variables, save for the **THEIL** index.[12]

Table 1 About Here

Given the relative time invariance of both the dependent and independent variables, the use of a regional-fixed effects estimator aiming to account for possible unobserved heterogeneity is ruled out. Beck (2001: p. 285) describes the problems by noting out that "with slowly changing independent variables, the fixed effect will soak up most of the explanatory power of those slowly changing variables". This, in turn, will most probably render these variables to be statistically insignificant.

---

[12] In addition, other variables used in the model also have some time variability (e.g. unemployment rate, GDP growth, related variety, unrelated variety, Hoover index and the local share of services) after an application of a variance decomposition analysis.

The aforementioned concerns necessitate the use of pooled ordinary least squares with heteroscedasticity robust standard errors. Thus, the basic pooled OLS model is expressed as:

$$PATINT_{rt} = a + \beta_1 RDEXP_{rt-1} + \beta_2 RDPERS_{rt-1} + \beta_3 RDGOV_{rt-1} +$$
$$\beta_4 RDHIGHED_{rt-1} + \beta_5 EMPSCITECH_{rt-1} + \beta_6 SCIENGIN_{rt-1} +$$
$$\beta_7 HTMANSERV_{rt-1} + \beta_8 HTMAN_{rt-1} + \gamma_1 HOOVIND_{rt-1} + \gamma_2 THEIL_{rt-1} +$$
$$\gamma_3 LOCALSERV_{rt-1} + \gamma_4 SIZE_{rt-1} + \gamma_5 POPDENS_{rt-1} + \gamma_6 GDPGR_{rt-1} +$$
$$\gamma_7 UNEMP_{rt-4} + e_{rt}, \tag{39}$$

where $PATINT$ is the dependent variable, while the variables in the right-hand side of the above equation are the independent and control variables of the model. In turn, $a$ is the constant term and $\beta = (\beta_1, \ldots, \beta_8)$ and $\gamma = (\gamma_1, \ldots, \gamma_7)$ are the vectors of coefficients of independent and control variables, respectively. On the other hand, $r$ and $t$ are the indices for regions ($r = 1, 2, \ldots, 13$) and time ($t = 1, 2, \ldots, 9$), while $e$ is the error term. The model also introduces independent and control variables with one lag[13].

Table 2 displays the descriptive statistics for all the variables of the model, while Table 3 presents the corresponding correlation coefficient matrix. The high correlation coefficients between different R&D spillover related variable and between population density and knowledge spillover variables stand out. This is to be expected since knowledge spillovers make up a significant part of agglomeration economies (Krugman 1991; see also Crescenzi et al. 2007).

---

[13] In the basic model, the unemployment rate enters with four lags. The present paper also uses unemployment rate with one lag to further check the effect of this variable on regional innovation.

Tables 2 and 3 About Here

Given some high correlation coefficients and in order to avoid multicollinearity issues, VIF factors were used as guidance. The average values of VIF corresponding to all model permutations are presented in the columns of Table 4. In all cases the average VIF values are kept well below the alarming levels.

The results of the pooled OLS estimation are presented in Table 4.

Table 4 About Here

In all the model permutations estimated, the R&D spillovers variables have a positive and significant effect on Greek regional patent intensity. This extends to both R&D expenditure and skilled labor variables. These results are in line with previous studies finding positive effects for both industrial and public R&D on regional innovation (Anselin et al. 1997; Buesa et al. 2010; Varga 2000). Similarly, the positive results of skilled-labor related indicators accord with earlier findings in a different country context suggesting that high technology and skilled labor are significant R&D spillovers factors fostering regional innovation (Audretsch and Feldman 1996). Once the results obtained on the other variables are commented upon, the discussion will return to the importance of the effect sizes relating to knowledge-spillovers variables.

Where the effect of the other explanatory variables is concerned, the results presented in Table 4 suggest that regional economic diversity (***THEIL*** index) and both related and unrelated variety all have a positive effect on regional innovation. These results accord with Feldman and Audretsch (1999) who highlight the advantages of diversity on regional innovation. In contrast, regional specialization (***HOOVER*** index) has a negative effect on regional innovation, a finding consistent with those by Feldman and Audretsch (1999) for U.S. MSAs and Crescenzi et al. (2007) for European regions and U.S. MSAs as well. Agglomeration economies[14], as measured by population density, stimulate regional innovation process, a result in line with the findings of Fritsch and Slavtchev (2010, 2011). The effect of the regional employment share in the services sectors is negative but insignificant. A negative result for this variable has been produced elsewhere (Fritsch and Slavtchev 2010, 2011). This seems quite reasonable given that it is the manufacturing sector that champions the patenting activity (see also Bode 2004; Bottazzi and Peri 2003).

On the other hand, regional GDP growth has a positive and significant effect on regional patent intensity, a finding consistent with earlier findings in the relative literature (Corradini and De Propris 2015; Moreno et al. 2005). This result is further complemented by the likewise positive effect of regional population growth. In contrast, the effect of regional unemployment rate has been negative and statistically insignificant in most of the cases. This effect is consistent with earlier findings (Crescenzi et al. 2007). The results on regional GDP and population growth, as well as that of unemployment, suggest that regional economic dynamism is a prerequisite for regional innovation. Finally, in the most permutations, the effect of firm size is negative but insignificant.

In order to assess the effect sizes of the knowledge-spillover related variables, a number of elasticities were calculated with the use of the corresponding estimated coefficients and the grand-means of the variables involved. These patent-intensity elasticities are presented in Table 5. What stands out is that the patent-elasticities in respect to the share of high-tech employment in manufacturing and services (0.59) and the scientists and engineers intensity (0.18) are much higher than the rest of the elasticities calculated. Thus, doubling the share of high-tech employment in manufacturing (100% increase) would increase the regional patent intensity by 59%, whereas doubling the intensity of scientists and engineers would increase patent intensity by 19%. These results put a strong emphasis on the importance of science and technology related regional human capital. Fritsch and Slavtchev (2010) report patent elasticities with respect to regional R&D employment (those having a tertiary degree and are employed as engineers or as natural scientists) that range from 0.4119 to 0.777.

---

[14] The other proxy of agglomeration economies which the present paper has tried, refers to the regional GDP as percentage of national GDP (Crescenzi et al. 2007). This variable also has positive and significant influences on regional innovation.

A number of previous studies have produced results emphasizing the relative importance of industrial (Autant-Bernard 2001; Jaffe 1989; Piergiovanni et al. 1997; Ronde and Hussler 2005) or university R&D spillovers (Acs et al. 1992; Blind and Grupp 1999; Del Barrio-Castro and Garcia-Quevedo 2005) on regional innovation.

The results obtained herein suggest that the patent elasticity in respect to university R&D is 0.126 and the corresponding one for manufacturing R&D 0.08. To put these elasticities into perspective, Anselin et al. (1997) have found that the estimated elasticity of innovation for university R&D is 0.11 for U.S. MSA's, while Fischer and Varga (2003) estimated elasticity values range from 0.13 to 0.21. On the other hand, in Pergionanni et al. (1997) the elasticity of innovation with respect to industrial R&D is 0.15, and in Piergiovanni and Santarelli (2001) the corresponding elasticity is 0.08. Fischer and Varga (2003) report innovation elasticities with respect to industrial R&D ranging from 0.10 to 0.40.

Table 5 About Here

Following Balli and Sørensen (2013), an orthogonalization process was applied to both the explanatory variables participating in an interaction. This essentially involved regressing one on another and taking the residuals. These residual terms were then used in the place of the original variables to construct the corresponding interaction term. This orthogonalization process takes place in order to reduce the multicollinearity-related problems. The results of the estimations involving interaction terms of orthogonalized variables are presented in Table 6 that follows.

Table 6 About Here

The results show that a significant number of interactions are developed in Greek regions. To an important extent, these interactions have a positive influence on regional patent intensity. Specifically, R&D personnel intensity in manufacturing seems to have a positive and significant effect on regional innovation when its interaction with the following variables is taken into account: a) scientists and engineers intensity, b) employment in science and technology intensity and c) the share of employment in high technology manufacturing[15]. These findings suggest that there could be knowledge spillovers facilitated through the interaction of manufacturing R&D personnel with highly qualified persons employed in a region that foster regional innovation. This can take place through their frequent communication, face-to-face contact as well as their participation in conferences and other seminars. There, there is the opportunity to exchange knowledge, ideas, advice and other information. In many circumstances, this exchange concerns differentiated ideas where knowledge spillover takes place among employees who belong to diversified industrial sectors (e.g. R&D manufacturing personnel and scientists and engineers or/and employees in science and technology). The specific findings concur with those of empirical studies in other country contexts (Charlot et al. 2015).

---

[15] The interaction of manufacturing R&D expenditure with the regional employment share of high technology manufacturing and high technology knowledge intensive services (**HTMANSERV**) is insignificant, whereas the corresponding interaction with regional employment share in hi-tech manufacturing (**HTMAN**) is positive and significant suggesting that the effect is mainly owned to the presence of hi-tech manufacturing.

The interaction between manufacturing R&D expenditures and higher education R&D expenditures is negative and of marginal statistical significance. This offers some tentative evidence suggesting for the possibility of some crowding out effect. A well-known fact is that in Greece the business sector R&D expenditure lags behind that of the public sector (Eurostat 2017). This needs to be considered when interpreting these results.

However, the interaction of manufacturing R&D personnel and R&D expenditure in higher education is positive and significant. The same applies to the interaction between manufacturing R&D personnel and the R&D expenditure in the government sector, but less statistical confidence can be placed on this result. These results suggest that manufacturing R&D personnel can benefit from public R&D efforts and that people are the key conduit facilitating cross-sector knowledge spillovers and synergies between the business and the public sector. These findings are in line with those of earlier studies that underline the benefits of collaboration between industrial R&D and public R&D expressed by university and government R&D (Fritsch and Franke 2004; Ponds et al. 2010).

## 6. Conclusions

This study attempts to fill part of the void pertaining to the lack of evidence when it comes to the determinants of regional innovation in Greece. The approach taken goes beyond a regional knowledge production function and draws possible explanations from a more extensive pool of R&D related variables and regional structural variables. Particular attention was paid to the role of employment related variables motivated by an understanding that the role of highly-qualified people is instrumental in facilitating knowledge spillovers and enhancing synergies between the business and public sector within the context of promoting regional innovation. In order to explore possible synergy effects, a number of interactions between key R&D related variables (both expenditure and employment related) as well as qualified-labour related variables were examined. The findings obtained contribute to a rather limited literature on the effects of such synergies coming mainly from a context of larger and more advanced economies. The paper also assesses the differences in the effect size of different sector-related R&D spillovers measures and compares these sizes to others in different regional contexts in USA and Europe. It is an interesting point of the present research as it shows where the different actors should focus regarding R&D. However, the novelty of the current research and differentiating factor compared to other studies refers to the fact that it employs game theory techniques in order to describe the licensing of the patents through royalties and to derive the subgame perfect Nash equilibrium under a Stackelberg duopoly.

The results obtained suggest that the R&D variables related to high-tech employment and R&D in manufacturing (expenditures and personnel) as well as R&D expenditures in public sector (higher education and government sector) have a positive and significant effect on regional patent intensity. The elasticities of regional innovation with respect to R&D expenditure is higher for the higher education and government sector rather than the manufacturing sector. Regional innovation elasticities in respect to employment measures, however, are even higher. The highest of these relates to the regional employment share in high-tech sectors followed by that regarding regional employment intensity in scientists and engineers.

These results highlighting the importance of people in the process of regional innovation are further corroborated by the results from the interactions explored. In these interactions the most significant synergy effects involve R&D personnel in manufacturing and other measures of highly-qualified (such as science and engineers) regional employment. There is evidence for positive synergies between the public and the private sector involving the R&D expenditure of the public sector and employment in manufacturing business R&D. This result is encouraging and the collaboration between the public sector (especially that of higher education) and the business sector should be the focus of policy efforts. This said, there is some concern regarding the possibility of crowding-out effects when it comes to the R&D expenditure at the regional level. Public R&D expenditure continues to be greater than that of the business sector and, at the regional level, some tentative evidence was produced that the individual positive effects of manufacturing and higher education R&D expenditure are more important than their interaction. This effect of expenditure imbalance is something that needs to attract the attention of the public debate and policy.

Regarding other possible determinants of regional innovation, the evidence produced favours the role of regional economic diversity, as opposed to regional specialization, and the presence of small firm structures. Regional innovation also seems to benefit from better regional economic conditions (higher GDP per capital growth and lower unemployment).

To sum up, the key findings and contributions of this paper highlight the importance of people in the facilitation of knowledge spillovers and fostering regional innovation.

There is also evidence that suggests synergies between the public and business sector. Again, the involvement of personnel in these interactions is instrumental. The key policy implication of the paper is that the emphasis should not only be placed on R&D expenditures but also on enhancing the regional capacities of highly qualified personnel across sectors.

Table 1. Variance Decomposition (N=117)

| Variable | BETWEEN-REGIONS/$\sigma^2$ | BETWEEN-TIME/$\sigma^2$ | RESIDUAL/$\sigma^2$ | SYSTEMATIC(MODEL)/$\sigma^2$ | F-REGION | F-TIME |
|---|---|---|---|---|---|---|
| **PATINT** | 0.644 | 0.04 | 0.316 | 0.896 | 16.27 (0.0) | 1.52 (0.16) |
| **RDEXP** | 0.959 | 0.003 | 0.038 | 0.962 | 201.3 (0.0) | 0.93 (0.5) |
| **RDPERS** | 0.851 | 0.012 | 0.137 | 0.863 | 49.84 (0.0) | 1.05 (0.41) |
| **RDGOV** | 0.865 | 0.019 | 0.116 | 0.884 | 59.84 (0.0) | 2.03 (0.05) |
| **RDHIGHED** | 0.979 | 0.004 | 0.016 | 0.984 | 478.3 (0.0) | 3.1 (0.003) |
| **HTMANSERV** | 0.917 | 0.009 | 0.074 | 0.926 | 99.68 (0.0) | 1.49 (0.17) |
| **HTMAN** | 0.893 | 0.016 | 0.091 | 0.909 | 78.19 (0.0) | 2.12 (0.04) |
| **SIZE** | 0.869 | 0.015 | 0.116 | 0.884 | 59.94 (0.0) | 1.58 (0.14) |
| **SCIENGIN** | 0.987 | 0.002 | 0.011 | 0.989 | 723.6 (0.0) | 1.98 (0.06) |
| **EMPSCITECH** | 0.993 | 0.001 | 0.006 | 0.994 | 1278 (0.0) | 2.05 (0.05) |
| **THEIL** | 0.595 | 0.354 | 0.051 | 0.949 | 93.05 (0.0) | 82.97 (0.0) |

Notes: F-test for regional and time effects (Probability of F in parentheses)

Table 2. Descriptive statistics of variables (N=117)

| Variables | Obs | Mean | Std. Dev. | Min | Max |
|---|---|---|---|---|---|
| 1. *PATINT* | 117 | 4.3633 | 4.5519 | 0.0043 | 16.9 |
| 2. *RDEXP* | 117 | 1.084 | 3.5947 | 0.0001 | 16.47 |
| 3. *RDPERS* | 117 | 0.0183 | 0.0615 | 8.1e-06 | 0.3105 |
| 4. *RDGOV* | 117 | 3.8726 | 10.3023 | 0.2834 | 68.7139 |
| 5. *RDHIGHED* | 117 | 6.1999 | 12.9603 | 0.7591 | 61.1739 |
| 6. *SCIENGIN* | 117 | 2.6044 | 6.8152 | 0.1 | 28.8866 |
| 7. *EMPSCITECH* | 117 | 13.3873 | 33.3109 | 1.2605 | 138.5242 |
| 8. *HTMANSERV* | 117 | 0.8968 | 1.0052 | 0.0001 | 4.4 |
| 9. *HTHMAN* | 117 | 0.9951 | 1.1088 | 0.0001 | 4 |
| 10. *UNEMP4* | 117 | 0.108 | 0.023 | 0.0685 | 0.18 |
| 11. *UNEMP1* | 117 | 0.1003 | 0.0231 | 0.0448 | 0.18 |
| 12. *GDPGR* | 117 | 0.0471 | 0.0383 | -0.0717 | 0.1505 |
| 13. *THEIL* | 117 | 4.2717 | 0.3178 | 3.7165 | 5.2268 |
| 14. *RELATED* | 117 | 0.839 | 0.2395 | 0.48 | 1.4919 |
| 15. *HOOVIND* | 117 | 33.9751 | 10.6009 | 18.034 | 70.738 |
| 16. *SIZE* | 117 | 5.2723 | 2.3555 | 1.4869 | 17.4704 |
| 17. *POPDENS* | 117 | 132.25 | 264.18 | 29.33 | 1051.3 |
| 18. *POPGR* | 117 | -0.0236 | 0.2175 | -0.3392 | 0.5487 |
| 19. *LOCALSERV* | 117 | 0.6035 | 0.0873 | 0.4529 | 0.7939 |
| 20. *UNRELATED* | 117 | 3.4326 | 0.1203 | 3.1591 | 3.7489 |

Table 3. Correlation matrix with correlation coefficients of variables (N=117)

| Variables | 1. | 2. | 3. | 4. | 5. | 6. | 7. | 8. | 9. | 10. | 11. | 12. | 13. | 14. | 15. | 16. | 17. | 18. | 19. | 20. |
|---|---|---|---|---|---|---|---|---|---|---|---|---|---|---|---|---|---|---|---|---|
| 1. *PATINT* | 1.00 | | | | | | | | | | | | | | | | | | | |
| 2. *RDEXP* | 0.55 | 1.00 | | | | | | | | | | | | | | | | | | |
| 3. *RDPERS* | 0.52 | 0.92 | 1.00 | | | | | | | | | | | | | | | | | |
| 4. *RDGOV* | 0.55 | 0.89 | 0.75 | 1.00 | | | | | | | | | | | | | | | | |
| 5. *RDHIGHED* | 0.60 | 0.96 | 0.88 | 0.93 | 1.00 | | | | | | | | | | | | | | | |
| 6. *SCIENGIN* | 0.58 | 0.97 | 0.88 | 0.95 | 0.99 | 1.00 | | | | | | | | | | | | | | |
| 7. *HMANSERV* | 0.7 | 0.8 | 0.72 | 0.77 | 0.81 | 0.82 | 1.00 | | | | | | | | | | | | | |
| 8. *HTMAN* | 0.5 | 0.59 | 0.58 | 0.55 | 0.6 | 0.61 | 0.75 | 1.00 | | | | | | | | | | | | |
| 9. *EMPSCITECH* | 0.58 | 0.97 | 0.89 | 0.95 | 0.99 | 0.99 | 0.82 | 0.6 | 1.00 | | | | | | | | | | | |
| 10. *UNEMP4* | -0.17 | -0.1 | -0.00 | -0.13 | -0.1 | -0.1 | -0.2 | -0.02 | -0.1 | 1.00 | | | | | | | | | | |
| 11. *UNEMP1* | -0.29 | -0.2 | -0.1 | -0.22 | -0.2 | -0.2 | -0.3 | -0.1 | -0.2 | 0.63 | 1.00 | | | | | | | | | |
| 12. *THEIL* | 0.44 | 0.54 | 0.46 | 0.6 | 0.54 | 0.56 | 0.51 | 0.52 | 0.56 | 0.03 | -0.2 | 1.00 | | | | | | | | |
| 13. *RELATED* | 0.46 | 0.56 | 0.5 | 0.59 | 0.56 | 0.57 | 0.6 | 0.67 | 0.57 | -0.0 | -0.2 | 0.95 | 1.00 | | | | | | | |
| 14. *HOOVIND* | -0.4 | -0.4 | -0.36 | -0.37 | -0.4 | -0.39 | -0.6 | -0.26 | -0.4 | 0.32 | 0.43 | -0.2 | -0.3 | 1.00 | | | | | | |
| 15. *SIZE* | 0.18 | 0.23 | 0.2 | 0.23 | 0.21 | 0.23 | 0.41 | 0.75 | 0.22 | 0.02 | -0.0 | 0.43 | 0.59 | -0.1 | 1.00 | | | | | |
| 16. *POPDENS* | 0.59 | 0.98 | 0.92 | 0.93 | 0.99 | 0.99 | 0.81 | 0.6 | 0.99 | -0.1 | -0.2 | 0.54 | 0.56 | -0.4 | 0.2 | 1.00 | | | | |
| 17. *POPGR* | 0.5 | 0.32 | 0.31 | 0.31 | 0.33 | 0.33 | 0.33 | 0.17 | 0.33 | -0.2 | -0.2 | 0.44 | 0.41 | -0.2 | -0.0 | 0.35 | 1.00 | | | |
| 18. *LOCALSERV* | 0.36 | 0.52 | 0.48 | 0.53 | 0.56 | 0.55 | 0.29 | 0.07 | 0.56 | -0.3 | -0.4 | 0.42 | 0.33 | -0.2 | -0.2 | 0.57 | 0.54 | 1.00 | | |
| 19. *UNRELATED* | 0.24 | 0.32 | 0.23 | 0.41 | 0.32 | 0.35 | 0.16 | 0.03 | 0.34 | 0.15 | -0.0 | 0.76 | 0.51 | 0.07 | -0.0 | 0.32 | 0.34 | 0.46 | 1.00 | |
| 20. *GDPGR* | 0.2 | 0.08 | 0.11 | 0.08 | 0.11 | 0.09 | 0.1 | 0.07 | 0.1 | 0.01 | -0.0 | -0.2 | -0.2 | 0.06 | -0.1 | 0.1 | -0.0 | 0.06 | -0.2 | 1.00 |

Table 4. Pooled OLS estimations (N=117) Dependent Variable: Patent intensity

| Variables | 1. | 2. | 3. | 4. | 5. | 6. | 7. | 8. | 9. | 10. | 11. |
|---|---|---|---|---|---|---|---|---|---|---|---|
| RDEXP | 0.3407*** (0.1191) | 0.3322*** (0.1182) | - | - | - | - | - | - | - | - | - |
| RDPERS | - | - | 20.0364*** (5.8308) | - | - | - | - | - | - | - | - |
| RDGOV | - | - | - | - | - | - | - | - | - | 0.1155** (0.0538) | - |
| RDHIGHED | - | - | - | - | - | - | - | 0.1236*** (0.0319) | 0.1235*** (0.0312) | - | - |
| SCIENGIN | - | - | - | 0.2573*** (0.0658) | - | - | - | - | - | - | - |
| HTMANSERV | - | - | - | - | - | 2.8559*** (0.4901) | - | - | - | - | - |
| EMPSCITECH | - | - | - | - | - | - | 0.0616*** (0.0123) | - | - | - | - |
| HTMAN | - | - | - | - | 2.1185*** (0.4732) | - | - | - | - | - | - |
| UNEMP4 | -12.764 (15.7925) | -7.6459 (15.5622) | - | -11.4445 (16.4264) | 0.043 (15.1068) | 5.3238 (14.6243) | -3.2049 (13.387) | -10.3097 (15.445) | -5.9816 (15.15) | -5.7274 (16.9729) | -12.3735 (15.4893) |
| UNEMP1 | - | - | -20.7649 (19.1654) | - | - | - | - | - | - | - | - |
| GDPGR | 29.8695*** (9.1684) | 28.535*** (8.8477) | 28.1871*** (8.8121) | 24.1864*** (8.9976) | 21.9222*** (8.6939) | 20.6233*** (8.1674) | - | 27.6409*** (9.0019) | 26.4061*** (8.6402) | 26.9278*** (9.1468) | 27.674*** (9.0612) |
| THEIL | 4.4317*** (1.4038) | - | 4.532*** (1.3635) | - | - | 2.4591* (1.457) | - | 3.9385*** (1.4384) | - | - | 4.0962*** (1.4311) |
| RELATED | - | 6.1771*** (2.1878) | - | - | 4.1549* (2.3654) | - | - | - | 5.4915*** (2.1901) | - | - |
| UNRELATED | - | - | - | 6.5875*** (3.3643) | - | - | - | - | - | 6.3205* (3.4985) | - |
| HOOVIND | -0.1033*** (0.0264) | -0.0935*** (0.0264) | -0.0914*** (0.0281) | -0.1066*** (0.0277) | -0.1027*** (0.0252) | -0.0212 (0.0275) | -0.0846*** (0.0304) | -0.0946*** (0.026) | -0.0853*** (0.0259) | -0.1217*** (0.0283) | -0.0954*** (0.0262) |
| SIZE | -0.1014 (0.1589) | -0.225 (0.1931) | -0.1137 (0.1657) | 0.1558 (0.135) | -0.7667*** (0.2305) | -0.3326** (0.1623) | 0.0428 (0.1322) | -0.1101 (0.1567) | -0.224 (0.1908) | 0.2436* (0.1358) | -0.1177 (0.1575) |
| POPDENS | - | - | - | - | - | - | - | - | - | - | 0.0059*** |

|  | (1) | (2) | (3) | (4) | (5) | (6) | (7) | (8) | (9) | (10) | (11) |
|---|---|---|---|---|---|---|---|---|---|---|---|
|  |  |  |  |  |  |  | (0.0016) |  |  |  |  |
| POPGR | - | - | - | - | - | - | 7.6909*** | - | - | - | - |
|  |  |  |  |  |  |  | (1.8093) |  |  |  |  |
| LOCALSERV | 1.0486 | 2.4413 | 0.1079 | 1.6977 | 5.8797 | 5.0223 | -5.808 | -0.8156 | 0.2446 | 6.0832 | -1.267 |
|  | (5.5622) | (5.5336) | (5.5411) | (5.3588) | (4.9986) | (4.9811) | (4.9692) | (5.619) | (5.6163) | (5.2666) | (5.7849) |
| CONS | -11.2391** | 1.4534 | -10.6889** | -16.7641 | 1.9046 | -10.6731** | 10.3386*** | -8.8456 | 2.5579 | -18.9353* | -8.9786 |
|  | (5.6956) | (3.9977) | (5.512) | (10.5274) | (3.6509) | (4.7289) | (3.9627) | (5.6629) | (3.8847) | (10.9673) | (5.7669) |
| $R^2$ | 0.4347 | 0.4284 | 0.4458 | 0.4412 | 0.4755 | 0.5477 | 0.4933 | 0.4551 | 0.4509 | 0.4080 | 0.4524 |
| F | 18.87 | 19.82 | 22.23 | 23.40 | 31.46 | 28.82 | 27.78 | 28.77 | 32.73 | 12.73 | 26.46 |
| Avg VIF | 1.83 | 2.05 | 1.85 | 1.65 | 2.37 | 1.94 | 1.66 | 1.9 | 2.1 | 1.69 | 1.9 |

\* significant at 10%, \*\* significant at 5%, \*\*\*significant at 1%

Table 5. Estimated patent intensity elasticities for R&D spillovers measures

| Variables | Elasticities |
|---|---|
| 1. R&D higher education intensity | 0.126 |
| 2. R&D government sector intensity | 0.107 |
| 3. R&D manufacturing intensity | 0.08 |
| 4. Share of high-tech employment in manufacturing and services | 0.59 |
| 5. Scientists and engineers intensity | 0.18 |

Table 6. Pooled OLS estimations with interaction effects (N=117) Dependent Variable: Patent intensity

| Variables | 1. | 2. | 3. | 4. | 5. | 6. | 7. | 8. | 9. | 10 | 11. |
|---|---|---|---|---|---|---|---|---|---|---|---|
| RDEXP*SCIENGIN | -0.0183 (0.0588) | - | - | - | - | - | - | - | - | - | - |
| RDEXP*EMPSCITECH | - | -0.0029 (0.0119) | - | - | - | - | - | - | - | - | - |
| RDEXP*HTMANSERV | - | - | -0.2258 (0.1985) | - | - | - | - | - | - | - | - |
| RDEXP*HTHMAN | - | - | - | 0.1879*** (0.0648) | - | - | - | - | - | - | - |
| RDPERS*SCIENGIN | - | - | - | - | 2.4417*** (0.8045) | - | - | - | - | - | - |
| RDPERS*EMPSCITECH | - | - | - | - | - | 0.687*** (0.2382) | - | - | - | - | - |
| RDEXP*RDHIGHED | - | - | - | - | - | - | -0.0405* (0.0222) | - | - | - | - |
| RDEXP*RDGOV | - | - | - | - | - | - | - | 0.0085 (0.0101) | - | - | - |
| RDPERS*RDHIGHED | - | - | - | - | - | - | - | - | 0.6415** (0.3006) | - | - |
| RDPERS*RDGOV | - | - | - | - | - | - | - | - | - | - | 0.5722* (0.3138) |
| RDPERS*HTECHMAN | - | - | - | - | - | - | - | - | - | 9.7281* (5.333) | - |
| RDEXP | -0.2152 (0.2332) | -0.3913* (0.212) | -0.2934** (0.1501) | 0.0655 (0.1186) | - | - | -0.476*** (0.1839) | 0.1908 (0.1355) | - | - | - |
| SCIENGIN | 0.3509*** (0.122) | - | - | - | 0.3749*** (0.1151) | - | - | - | - | - | - |
| EMPSCITECH | - | 0.0855*** (0.0262) | - | - | - | 0.094*** (0.0294) | - | - | - | - | - |
| HTMANSERV | - | - | 3.7128*** (0.6099) | - | - | - | - | - | - | - | - |
| HTMAN | - | - | - | 2.1299*** | - | - | - | - | - | 2.1178*** | - |

|  | (1) | (2) | (3) | (4) | (5) | (6) | (7) | (8) | (9) | (10) | (11) |
|---|---|---|---|---|---|---|---|---|---|---|---|
|  |  |  |  | (0.5419) |  |  |  |  |  | (0.612) |  |
| *RDPERS* | - | - | - | - | -2.3613 (7.8594) | -9.3122 (9.539) | - | - | -0.2652 (6.5107) | 6.8078 (6.3974) | 12.5082** (6.2011) |
| *RDHIGHED* | - | - | - | - | - | - | 0.2129*** (0.0499) | - | 0.1566*** (0.0446) | - | - |
| *RDGOV* | - | - | - | - | - | - | - | 0.0833 (0.0904) | - | - | 0.125 (0.0783) |
| *THEIL* | - | - | - | 3.4605*** (1.4315) | 4.0949*** (1.4461) | 4.148*** (1.4421) | 3.9193*** (1.4454) | - | - | - | 4.1812*** (1.4786) |
| *RELATED* | - | 5.4203*** (2.2303) | - | - | - | - | - | 5.9254*** (2.2406) | 5.4431*** (2.2075) | 3.6818 (2.2871) | - |
| *UNRELATED* | 6.2554* (3.4533) | - | 5.1925* (3.1227) | - | - | - | - | - | - | - | - |
| *UNEMP4* | -9.5372 (17.3932) | -3.568 (15.8319) | 10.824 (16.0343) | -5.9221 (15.4581) | -11.2831 (16.063) | -10.998 (16.0944) | -9.0779 (15.9981) | -5.0541 (16.0374) | -5.9457 (15.8947) | -2.0975 (15.8239) | -13.6491 (16.1964) |
| *GDPGR* | 23.5827*** (9.0114) | 26.252*** (8.6906) | 18.647*** (7.4771) | 22.122*** (8.8347) | 27.104*** (8.947) | 27.13*** (8.986) | 25.9543*** (9.0384) | 28.818*** (8.892) | 26.307*** (8.6404) | 20.125** (8.7144) | 28.2914*** (9.0894) |
| *HOOVIND* | -0.1066*** (0.0281) | -0.0893*** (0.0264) | -0.0188 (0.0285) | -0.0829*** (0.0261) | -0.0934*** (0.0265) | -0.095*** (0.0265) | -0.0938*** (0.0263) | -0.0934*** (0.0266) | -0.0832*** (0.0265) | -0.075*** (0.0273) | -0.0949*** (0.0268) |
| *SIZE* | 0.1619 (0.136) | -0.203 (0.1919) | -0.1547 (0.1529) | -0.5863*** (0.2243) | -0.1118 (0.1569) | -0.1009 (0.1573) | -0.0887 (0.1579) | -0.2199 (0.1934) | -0.2133 (0.1922) | -0.6304*** (0.2377) | -0.1119 (0.1588) |
| *LOCALSERV* | 2.0875 (5.4723) | 0.8076 (5.6584) | 10.4175** (5.4135) | 2.8986 (5.8981) | -0.7454 (5.6266) | -1.1088 (5.6843) | -0.8684 (5.6867) | 2.1089 (5.6338) | 0.1363 (5.6829) | 4.4234 (5.6378) | -0.5137 (5.675) |
| CONS | -16.1112 (10.6953) | 2.0906 (4.0024) | -23.446*** (8.8283) | -8.3604 (5.3808) | -9.3699 (5.8131) | -9.5211* (5.7972) | -9.1182 (5.744) | 1.4355 (4.0343) | 2.4483 (3.9933) | 1.9439 (3.8665) | -9.5291 (5.9728) |
| $R^2$ | 0.4435 | 0.4512 | 0.5626 | 0.4946 | 0.4654 | 0.4642 | 0.4656 | 0.4327 | 0.4555 | 0.4839 | 0.4515 |
| F | 19.11 | 23.93 | 26.62 | 37.81 | 51.40 | 31.89 | 39.49 | 75.67 | 28.98 | 31.14 | 22.96 |

\* significant at 10%, ** significant at 5%, ***significant at 1%